\documentclass[prl,twocolumn,superscriptaddress,showpacs,floatfix]{revtex4}
\usepackage{graphics}
\usepackage{graphicx}
\begin{document}
\newcommand{\RR}{\mathrm{\mathbf{R}}}
\newcommand{\rr}{\mathrm{\mathbf{r}}}
\newcommand{\defin}{\stackrel{def}{=}}

\title{Quantum control of donor electrons at the Si-SiO$_2$ interface}
\author{M.J. Calder\'on}

\affiliation{Condensed Matter Theory Center, Department of Physics,
University of Maryland, College Park, MD 20742-4111}

\author{Belita Koiller}

\affiliation{Condensed Matter Theory Center, Department of Physics,
University of Maryland, College Park, MD 20742-4111}

\affiliation{Instituto de F\'{\i}sica, Universidade Federal do Rio de
Janeiro, Caixa Postal 68528, 21941-972 Rio de Janeiro, Brazil}

\author{Xuedong Hu}

\affiliation{Department of Physics, University at Buffalo, SUNY, Buffalo, NY 14260-1500}

\author{S. Das Sarma}

\affiliation{Condensed Matter Theory Center, Department of Physics,
University of Maryland, College Park, MD 20742-4111}

\date{\today}



\begin{abstract}

Prospects for the quantum control of electrons in the silicon 
quantum computer architecture are considered theoretically. In particular, 
we investigate the feasibility of shuttling donor-bound electrons between 
the impurity in the bulk and the Si-SiO$_2$ interface by tuning an
external electric field. We calculate the shuttling time to range 
from sub-picoseconds to nanoseconds depending on the distance 
($\sim 10-50$ nm) of the donor from the interface. Our results 
establish that quantum control in such nanostructure architectures
could, in principle, be 
achieved.

\end{abstract}

\maketitle
\pagebreak

Silicon based structures are among the most promising candidates 
for the development of a
quantum computer~\cite{Kane,Vrijen,skinner03,Barrett,martin03,friesen04,Kane_MRS} 
due to the existing 
high level of nanofabrication control as well as the well-established scalability advantages of Si microelectronics.  
Different architectures have been proposed in which 
nuclear spin~\cite{Kane,skinner03}, 
electron spin~\cite{Vrijen} or electron charge~\cite{Barrett} 
are used as qubits. 
All of these proposals ultimately rely on the quantum control of 
single electrons 
bound to donors. The ability to move or `shuttle' electrons between 
a donor and the Si surface using external electric field is an 
essential element of Si quantum computer architectures because the 
measurement of the electron spin states can only occur at surfaces whereas the qubit 
entanglement takes place at the donor sites. In particular, this
shuttling time must be much shorter than the spin dephasing time ($\sim
1$ ms in bulk Si). 
Application of electrostatic potentials at surface electrodes 
would drag the electron from and to the donor allowing the manipulation of the 
electron-donor coupling. Quantum control of donor states is also a
crucial consideration in Si device `roadmap' as miniaturization of transistors leads to only a few dopants
per device.

In this Letter we theoretically consider the problem of quantum control of donor-bound electrons near a Si-SiO$_2$ interface. We investigate the precise 
extent to which a donor bound electron in the bulk (a few tens of nanometers 
from the interface) can be manipulated between the donor-bound 
state and a surface-bound state (within a few nanometers of the
interface) by suitably tuning an external electric field applied perpendicular 
to the interface. We address two issues of 
paramount importance in this context: (i) How fast can this electron 
shuttling between the donor and the interface be done in realistic Si 
structures? (ii) Is the shuttled electron at the surface still localized 
in all three dimensions (which will allow to take it back to the donor) 
or is it a delocalized two-dimensional electron 
(which will make it impossible to measure its spin)? 
Our quantitative answers to these questions indicate that 
quantum control consistent with Si quantum computer architectures 
could, in principle, be achieved.  

We consider a single electron bound to a substitutional P donor in Si, a distance $d$ from an ideally flat Si/SiO$_2$ (001) interface, under an applied uniform electric field ${\bf F}$ perpendicular to the interface. Although conceptually simple, this one-electron problem has no formal solution. The approach adopted here is based on well established approximations\cite{Kohn,stern67,martin78,macmillen84}, proposed in the context of conventional Si-based devices, and validated by extensive studies available in the literature \cite{ando82}. The formalism, briefly outlined below, allows us to keep the essential physical aspects of the problem within a clear and realistic description. 
The Hamiltonian is written in the single-valley effective-mass approximation \cite{macmillen84}:
\begin{equation}
H = T +\kappa e F z -{{2}\over{r}}+{{2Q}\over{\sqrt{\rho^2+(z+2d)^2}}}-{{Q}\over{2(z+d)}}~,
\label{eq:h}
\end{equation} 
where $T = -\left({{\partial^2}\over{\partial
x^2}}+{{\partial^2}\over{\partial
y^2}}+\gamma{{\partial^2}\over{\partial z^2}}\right)$,  $\gamma=
m_\perp/m_\|$ is the ratio between the transverse ($m_\perp = 0.191 \,
m$) and longitudinal ($m_\|=0.916\,m$) effective masses, accounting for the Si conduction band valley's anisotropy, and ${\vec\rho} = (x,y)$. 
This equation refers to one of the band minima along $z$, which become lower 
in energy for this geometry \cite{ando82}.
Lengths and energies are given in rescaled atomic units: $a^*={{\hbar^2\epsilon_1}/{m_\perp e^2}} = 3.157$ nm~ and $Ry^*={{m_\perp e^4}/{2\hbar^2\epsilon_1^2}}= 19.98$ meV respectively, $\kappa=3.89 \times 10^{-7} \epsilon_1^3 \left({{m}/{m_\perp}}\right)^2$ cm/kV, the electric field $F$ is given in kV/cm, and $Q={{(\epsilon_2-\epsilon_1)}/{(\epsilon_2+\epsilon_1)}}$, where $\epsilon_1=11.4$ and  $\epsilon_2=3.8$ are the Si and SiO$_2$ static dielectric constants. The second term in Eq.~(\ref{eq:h}) is the electric field linear potential, the third term is the donor attractive potential, and the last two terms are the donor and electron image potentials respectively. In this case $Q<0$, meaning that the images keep the same sign as their originating charges.

\begin{figure}
\begin{center}
\resizebox{85mm}{!}{\includegraphics{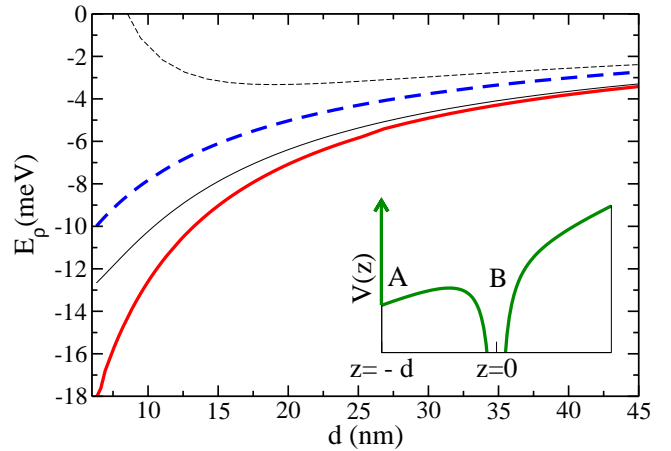}}
\caption{\label{fig:interface}(Color online) The broad lines, solid and dashed respectively, give the ground and first excited state energy terms ($E_\rho$ and $E'_\rho$) obtained variationally as a function of the donor-to-interface distance. The narrow lines give the same energies calculated within the parabolic approximation. The inset displays an outline of the double-well potential.}
\end{center}
\end{figure}

The overall potential profile for the donor electron along the $z$-axis  is schematically shown on the inset in Fig.~\ref{fig:interface}. Note that this is equivalent to an asymmetric  double-well configuration: The well near the interface is denoted by $A$ and the one around the donor site by $B$. 
Assuming the wells are not coupled, we obtain  variationally each well's ground state, defining a basis set $\{\Psi_A,\Psi_B\}$ for the 
low-lying energy eigenstates of the double-well. 
Truncating the Hilbert space into this particular two-dimensional subspace, based on the double-well analogy, is meaningful only for 
sufficiently large donor-interface separations. If $d\lesssim a^*$, a
single well description is more appropriate \cite{macmillen84}, and we
therefore limit the range of distances examined here to $d>2 a^*$. Moreover, each well's ground state must have an excitation gap to the 
first excited state that is much larger than $k_B T$, which sets an upper bound for the temperature as well as for $d$. For $d\to\infty$, 
the interface does not play a role and the single-valley approximation breaks down \cite{friesen05}. 
Uncoupled effective mass Hamiltonians for the donor electron in the $A$- and $B$-regions are written as $H_i=T+V_i$, $i=A$, $B$, with 
\begin{equation}
V_A=\kappa eFz-{{2}\over{\sqrt{\rho^2+d^2}}}+{{2Q}\over{\sqrt{\rho^2+d^2}}}-{{Q}\over{2(z+d)}},
\label{eq:va}
\end{equation}
and $V_B=-{{2}/{r}}$. The donor-related terms [third and fourth terms in Eq.~(\ref{eq:h})] are approximated in $V_A$ by their value at the interface: $V_{\rm P^+} \approx 2(-1+Q)/\sqrt{\rho^2+d^2}$, providing  confinement along $x$-$y$ in the $A$-region. Further assuming that $d\gg \rho$ leads to the 2-D parabolic potential approximation suggested in Ref.~\onlinecite{Kane00PRB}:  $V_{\rm parab}(\rho) = (1-Q)(-2/d + \rho^2/d^3)$. The barrier at the oxide interface is assumed to be infinite, so $\Psi_i=0$ for $z<-d$. The following properly normalized variational forms were adopted for $z>-d$ \cite{Kohn,macmillen84}:
\begin{eqnarray}
\Psi_A&=&f_\alpha(z)\times g_\beta(\rho)\nonumber\\ 
&=&  {{\alpha^{\frac{2\ell+1}{2}}}\over{\sqrt{(2\ell)!}}} (z+d)^{\ell}\, e^{-{\alpha (z+d)}/{2}}\times{{\beta}\over{\sqrt{\pi}}}\, e^{{-\beta^2 \rho^2}/{2}}
\label{eq:psia}\\
\Psi_B&\propto&(z+d)e^{-\sqrt{\rho^2 / a^2+z^2/b^2}}~,
\label{eq:psib}
\end{eqnarray}
where $\alpha,~\beta,~a$ and $b$ are variational parameters chosen to
minimize $E_i=\langle \Psi_i |H_i| \Psi_i\rangle$ for $i=A,~B$. In
$f_{\alpha}(z)$ we have used $\ell=2$ which gives better agreement
with the exact wave-function for the infinite triangular well
\cite{stern72} than $\ell=1$ or $\ell=3$, as well as a lower variational energy than $\ell=1$, and essentially the same as $\ell=3$.
For $d> 2a^*$, we find that $a$ and $b$ coincide with the
Kohn-Luttinger (KL) variational Bohr radii for the isolated impurity
($d\rightarrow \infty$), where $a=2.365$ nm and $b=1.36$ nm.

\begin{figure}
\resizebox{85mm}{!}{\includegraphics{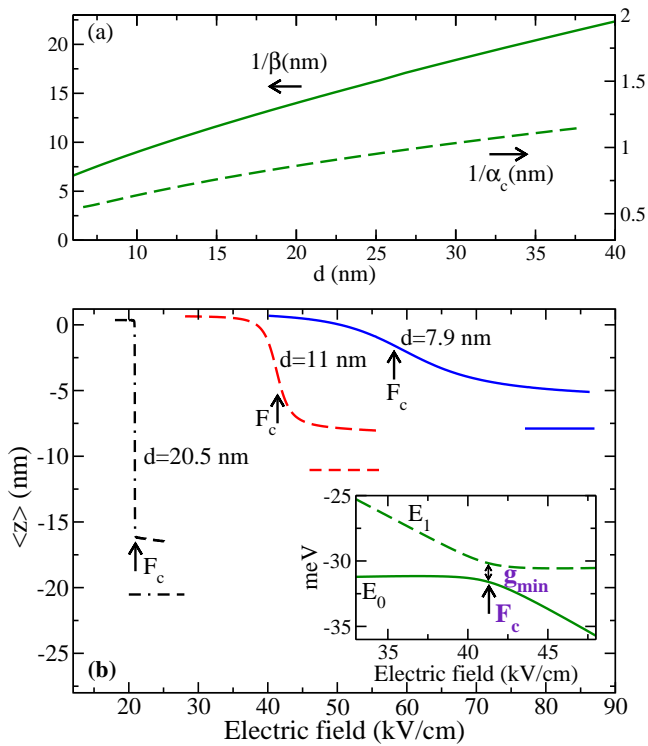}}
\caption{\label{fig:zav}(Color online)(a) Confinement lengths $1/\beta$ and $1/\alpha_c$ (for $\alpha$ calculated at $F=F_c$) obtained variationally. (b) Expectation value of the electron $z$-coordinate versus electric field intensity $F$ for three values of the donor distance to the barrier $d$. The horizontal lines indicate the barrier position in each case. The inset shows the eigenvalues E$_0$ and E$_1$ as a function of $F$ for $d=11$ nm. 
}
\end{figure}

A relevant question in the present context regards the electron
confinement parallel to the $xy$ plane when it is drawn towards the
interface ($A$-region) by the field. It still remains bound to the donor through $V_{\rm P^+}$. The calculated energy  as a function of $d$, $E_\rho=\langle g|H_A|g\rangle$, is plotted in Fig.~\ref{fig:interface}. The energy of the first excited state 
calculated variationally assuming the functional form $g'\propto x\,g_{\beta'}(\rho)$, 
$E'_\rho=\langle g'|H_A|g'\rangle$, is also shown. 
For $d=30$ nm, we get significant binding ($|E_\rho|\sim 5$ meV) for the ground state.
Results obtained within the parabolic approximation are also shown in Fig.~\ref{fig:interface}. As expected, this approximation underestimates the binding energies and overestimates the gap between successive levels, though convergence towards the variational results is obtained as $d$ increases. The variational parameter $\beta$, characterizing the radial confinement of the ground state parallel to the interface, is given in Fig.~\ref{fig:zav}(a). 
For the coherent manipulation of electrons in quantum devices, it is
crucial that the entanglement of the electronic states occurs in a 
completely controlled and reversible manner. This requires that the ionized state near the barrier remains laterally bound to its
respective donor site, setting an upper bound for the operating 
temperature [$k_B T\ll min(|E_\rho|,E'_\rho - E_\rho$)] 
as well as for the donor planar density [$n < (\beta/2)^2$] to 
avoid significant 
wavefunction overlap among electrons bound to neighboring donors.
For $d=30$ nm, we get an excitation gap $E'_{\rho}-E_\rho \sim 1$ meV, 
and donor electron wavefunction confinement within a $\sim 40$ nm 
diameter region parallel to the interface. These parameters yield an upper bound of $n \sim 10^{10}{\rm cm}^{-2}$ for 
planar donor densities and limit the operating temperature to a few K. Above these limits,
the electrons at the Si surface would either form a delocalized
impurity band ($n> 10^{10} {\rm cm}^{-2}$) or become thermally excited ($T>
1-5 K$).

The double-well problem is solved through direct diagonalization of Eq.~(\ref{eq:h}),
$H=T+V_A+V_B+{2}/{\sqrt{\rho^2+d^2}}$, in the non-orthogonal basis 
$\{\Psi_A,\Psi_B\}$, leading to the two lowest energy eigenstates 
$\Psi_0$ and $\Psi_1$ and eigenvalues $E_0$ and $E_1$. The last term
in the expression for $H$ is added to avoid double counting of the donor potential which is partially included in $V_A$ through 
$V_{P^+}$.
The response of the electron to an applied electric field is depicted in Fig.~\ref{fig:zav}(b), where the expectation value $\langle \Psi_0|z|\Psi_0 \rangle$ is given for three values of $d$. 
At very low fields the ground state is centered around the donor site $(\Psi_0 \approx \Psi_B)$, and its response to increasing fields is strongly dependent on $d$:  For the smaller $d$ values, the electron is smoothly drawn from near the P$^+$ nucleus toward the barrier as $F$ increases, while for the larger $d$ the transition is more abrupt, and takes place at lower values of $F$ \cite{smit03,martins04}. 
We define a critical field $F_c$ as the field value at which the gap is minimized, $(E_1-E_0)_{\rm min}=g_{\rm min}$, characterizing the anticrossing point in a ($E_0$, $E_1$) vs $F$ diagram [see inset in Fig.~\ref{fig:zav}(b)].  For stronger fields $(F > F_c)$, $\Psi_0$ approaches $\Psi_A$. 
In this regime, the variational parameter $\alpha$ characterizing the decay of the wavefunction along $z$ [see Eq.(~\ref{eq:psia})] becomes relevant. Fig.~\ref{fig:zav}(a) gives the length $1/\alpha_c$ vs $d$ for $\alpha$ calculated at $F_c$. We note that for the range of distances studied here, $1/\alpha_c$ is always smaller than $a^*$, justifying the double-well approach discussed above. For larger values of $F$, confinement along the field direction becomes even stronger, with $1/\alpha$ decreasing by about a factor of 3 as $F$ increases from $F_c$ to $10 F_c$.  
The field-independent confinement length parallel to the interface, $(1/\beta)$, is typically one order of magnitude larger than $1/\alpha_c$, and both increase sublinearly with increasing $d$. 
\begin{figure}
\begin{center}
\resizebox{85mm}{!}
{\includegraphics{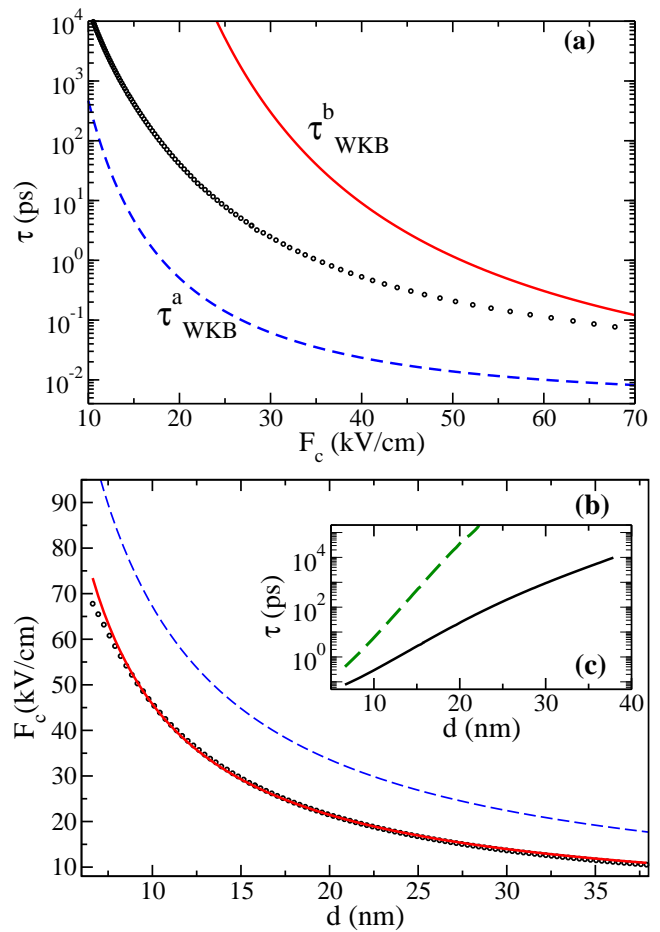}}
\caption{\label{fig:times}(Color online) (a) The data points give our
estimate for the characteristic donor ionization time, defined here as
$\tau = \hbar/g_{min}$, versus $F_c$. For comparison, we present the
inverse ionization rates obtained from the semiclassical approximation
(WKB) for the isotropic hydrogenic model with effective Bohr radii
equal to each of the variational KL parameters $a$ and $b$. (b)
Critical field versus distance of the donor to the interface. The
solid line gives a phenomenological fit $F_c \propto 1/|z_0|$. The
dashed line is a fitting to tight-binding results in
Ref.~\cite{martins04} of the form $F_c \propto 1/d$. (c) Donor
ionization tunneling (solid line) and adiabatic passage (dashed) times versus $d$.
}
\end{center}
\end{figure}
We now focus on a key parameter for general device applications: 
The tunneling time for donor ionization under an applied field. 
This question has traditionally been addressed in the
literature~\cite{banavar79} 
in analogy with the hydrogen atom ionization problem, 
based on the semiclassical WKB approximation. The expression 
for the tunneling time of an isotropic hydrogenic atom with 
effective Bohr radius $a_{\rm eff}$ in Si under a uniform electric 
field $F$ is~\cite{landau}
\begin{equation}
\tau_{\rm WKB}^{a_{\rm eff}}(F)=\frac{\epsilon_1 \hbar}{4e^3} a_{\rm eff}^3 F \exp\left({\frac{2e}{3\epsilon_1 a_{\rm eff} F}}\right)~.
\end{equation}
We present a fully quantum-mechanical estimate for the tunneling time, 
which we relate here to the anticrossing energy gap through the
uncertainty 
relation. Fig.~\ref{fig:times}(a) presents our results for $\tau =
\hbar/g_{\rm min}$ 
vs critical field, allowing direct comparison with the WKB estimates 
$\tau_{\rm WKB}^{a_{\rm eff}}(F_c)$. Using for $a_{\rm eff}$ the KL 
variational Bohr radii $a$ and $b$ leads, respectively, to lower and 
upper bounds for the calculated $\tau$, within a factor of up to 100  
in the range of electric fields considered 
here.  Our results cannot be fitted by the WKB isotropic model with an 
intermediate value of $a_{\rm eff}$, 
a result that may be due to the intrinsically anisotropic nature of
the system, 
to the final state in the ionized regime also being
a bound state here, as well as to limitations of the semiclassical approximation.  

A summary of our main results as a function of $d$, which is the single 
fabrication-related parameter in our model system, is also presented in 
Fig.~\ref{fig:times}. 
Fig.~\ref{fig:times}(b) gives the critical field 
versus $d$ results, which are well fitted by the function $F_c \propto
1/|z_0|$, where  $z_0=d-5/\alpha$ is the saturation value of $\langle
z \rangle$ after complete donor ionization. Results varying as $F_c \propto 1/d$, obtained through a tight-binding model where the Si conduction band details, in particular its six-fold degeneracy, are incorporated \cite{martins04}, are also given. 
The good qualitative agreement between the two curves indicates 
that our approach indeed captures the essential physical aspects of the 
system. The quantitative differences may arise not only from the 
single valley effective mass approximation adopted here, but also from the 
different geometries considered. 
These differences are
manifest in the fact that the relevant distances in the
phenomenological fittings are $|z_0|$ and $d$
respectively. Fig.~\ref{fig:times}(c) gives the tunneling times vs
$d$, showing that $\tau$ increases by 5 orders of magnitude as the donor
distance to the interface increases by a factor of 5, with $\tau$ ranging from
sub-picosecond to nanosecond time scales (e.g. $\tau \sim 3$ ps for
$d \simeq 15$ nm). Typical adiabatic passage times~\cite{martins04} $\tau_a= \hbar
|e| F_c d/ g_{\rm min}^2$, also shown, are orders of
magnitude larger than the tunneling time (e.g. $\tau_a \sim 0.4$ ns for $d \simeq 15$ nm).

The results presented here 
define critical parameters to be taken into account in a variety of scenarios where these processes are involved. 
In silicon-based nano-electronic devices~\cite{ono05}, the critical field dependence on donor positioning and the tunneling time provide 
relevant information concerning threshold-voltage control and device switching times respectively. 
For the reversible manipulation of electrons in quantum devices, 
properties of the states bound to the interface define 
upper bounds for the operating temperatures and for the donor planar
densities required 
($n<10^{10} {\rm cm}^{-2}$) to avoid significant 
wavefunction overlap among electrons bound to neighboring donors.
The times required for the shuttling processes will be different
depending on whether quantum information is stored in spin
or in charge degrees of freedom. 
In the case of spin qubits, tunneling times define the limiting time scales, since electron tunneling does not affect spin coherence. Spin
coherence times in bulk Si ($T_2 \sim 1$ms)~\cite{sousa03} are at
least 5 orders of magnitude longer than the tunneling times reported here.
Moreover, spin coherence in Si can be further enhanced by isotopic
purification~\cite{sousa03,tyryshkin03}. Recent experiments~\cite{schenkel05} demonstrate that $T_2$ is also sensitive to the
dopant depth below the interface as well as to the interface quality. This means that, for the particular geometry of interest here, the Si
bulk values of $T_2$ give an upper bound for the coherence times:
Careful interface optimization as well as avoiding inhomogeneities in
device fabrication and applied fields constitute additional requirements for a large number of operations to be performed before spin coherence is
lost. For charge qubits, orbital/charge coherence is required and adiabatic evolution of the electron state must take place, requiring much
longer time scales. Charge relaxation times are much shorter than spin coherence
times, of the order of $200$ns for Si quantum dots surrounded by oxide layers~\cite{gorman05}, but still longer than the adiabatic passage
times for $d<20$nm.
Our realistic results provide physical bounds for quantum control of qubits
based on Si:P donor electron states. In particular, provided that spin coherence times near a gated interface are reduced by no more than one
order of magnitude as compared to the bulk values
~\cite{foottime}, gate-voltage induced spin manipulation on the single
electron level may be feasible in donor-based Si quantum computer
architectures, similar to the exciting recent experimental results on
single electron manipulations in gated GaAs quantum dot nanostructures \cite{petta05}.

We thank B. Kane and K. Brown for calling our attention to this problem.
This work is supported by LPS and NSA. BK
acknowledges support by CNPq and 
FAPERJ.

\bibliography{control-v3}

\begin{thebibliography}{28}
\bibitem[{foo()}]{foottime}
\bibinfo{note}{Even shorter values of T$_2$ may be acceptable for experiments where spin-projection is the quantity of interest, and spin phase information is not relevant.}
\end{thebibliography}

\end{document}